\documentclass[final,5p,times]{elsarticle}
\pdfoutput=1
\usepackage[svgnames, table]{xcolor}
\usepackage{array, booktabs, makecell, multirow}
\usepackage{amsmath}
\usepackage{booktabs}
\usepackage{bigstrut}
\usepackage{float}
\usepackage{stfloats} 
\usepackage{booktabs}
\usepackage{graphicx}
\usepackage{amssymb}
\usepackage{multicol}
\usepackage[table]{xcolor}
\usepackage{multirow}
\usepackage[T1]{fontenc}
\usepackage{booktabs}
\graphicspath{ {./images/} }
\usepackage{url}
\begin{document}
	\begin{frontmatter}
		\date{}
		\title{Temporal Deep Learning Architecture for Prediction of COVID-19 Cases in India}
		\author[add1]{Hanuman Verma}
		\ead{hv4231@gmail.com}
		\author[add2]{Saurav Mandal}
		\ead{saurav.mnnit@gmail.com}
		\author[add3]{Akshansh Gupta }
		\ead{akshanshgupta@ceeri.res.in}
		
		\address[add1]{Bareilly College, Bareilly, Uttar Pradesh, India 243005}
		\address[add2]{School of Computational and Intergative Sciences, Jawaharlal Nehru University, New Delhi, India 110067}
		\address[add3]{CSIR-Central Electronics Engineering Research Institute,Pilani Rajasthan, India 333031}
		
		\begin{abstract}
			To combat the recent coronavirus disease 2019 (COVID-19), academician and clinician are in search of new approaches to predict the COVID-19 outbreak dynamic trends that may slow down or stop the pandemic. Epidemiological models like Susceptible-Infected-Recovered (SIR) and its variants are helpful to understand the dynamics trend of pandemic that may be used in decision making to optimize possible controls from the infectious disease. But these epidemiological models based on mathematical assumptions may not predict the real pandemic situation. Recently the new machine learning approaches are being used to understand the dynamic trend of COVID-19 spread. In this paper, we designed the recurrent and convolutional neural network models: vanilla LSTM, stacked LSTM, ED-LSTM, Bi-LSTM, CNN, and hybrid CNN+LSTM model to capture the complex trend of COVID-19 outbreak and perform the forecasting of COVID-19 daily confirmed cases of 7, 14, 21 days for India and its four most affected states (Maharashtra, Kerala, Karnataka, and Tamil Nadu). The root mean square error (RMSE) and mean absolute percentage error (MAPE) evaluation metric are computed on the testing data to demonstrate the relative performance of these models. The results show that the stacked LSTM and hybrid CNN+LSTM models perform best relative to other models.
		\end{abstract}
		\begin{keyword}
			Deep learning \sep COVID-19 \sep CNN \sep LSTM 
		\end{keyword}
	\end{frontmatter}

	\section{Introduction}\label{DL}
	The coronavirus disease 2019 (COVID-19) was identified in Wuhan city of China in December 2019 that arises due to severe acute respiratory syndrome coronavirus 2 (SARS-CoV-2) \cite{huang2020clinical}. It is categorized as an infectious disease and spreads among people through coming in close contact with infected people generally via small droplets due to coughing, sneezing, or talking, and through the infected surface. On March 11, 2020, the World Health Organization (WHO) declared the COVID-19 as a pandemic of infectious disease. In India, the first case of COVID-19 was reported in Kerala on January 30, 2020 and gradually spread throughout India especially in urban area, and India witnessed the first wave of COVID-19. India witnessed the second wave in March 2021, which was much more devastating than the first wave, with shortages of hospital beds, vaccines, oxygen cylinder and other medicines in parts of the country. To fight with the COVID-19, the country has vaccination, herd immunity, and epidemiological interventions as few possible options. In the early stage of COVID-19, India had imposed complete as well as partial lockdown as epidemiological interventions during the first wave that slowed the transmission rate and delayed the peak, and resulted in a lesser number of COVID-19 cases. India is the second most populous country in the world, where 68.84 \% and 31.16 \% India’s population lives in rural areas and urban areas respectively. The population density in northeast India is low in comparison to other states of India. The chance of getting infection depends on the spatial distance between the contacts and low-density population is less prone in comparison to high density population. Individual personal behavior (social distancing, frequent hand sanitation, and wearing a mask, etc.) also plays a key role to control the COVID-19 spread.
	
	Prediction of COVID-19 new cases per day will help the administration and planners to take the proper decision and help them in making effective policy to tackle the pandemic situation. The epidemiological models are very helpful to understand the trend of COVID-19 spread and useful in predicting the spread rate of the disease, the duration of the disease, and the peak of the infectious disease. It can be used for short term and long term predictions for new confirmed COVID-19 cases per day that may be used in decision making to optimize possible controls from the infectious disease. In literature, several mathematical models for infectious diseases such Logistic models \cite{turner1976theory}, generalized growth models \cite{chowell2017fitting}, Richards’s models \cite{richards1959flexible}, sub epidemics wave models \cite{chowell2019novel}, Susceptible-Infected-Recovered (SIR) model \cite{kermack1927contribution}, and Susceptible-Exposed-Infectious-Removed (SEIR) have been introduced. The SIR model is a compartmental model that considers the whole population as a closed population and divides this closed population into susceptible, infected, and recovered compartments. Few infected persons infect some other persons at an average rate $ R0 $, known as the basic reproduction number. Recently, some works have been reported in the literature using the SIR and its variants model to predict the COVID-19 outbreak \cite{chen2020time, cooper2020sir, bagal2020estimating, ardabili2020covid, verma2020analysis}. These epidemiological models are good in understanding the trend of COVID-19 spread but are designed based on several assumptions that would not hold generally on real-life data\cite{chimmula2020time}. It is unreliable due to the complex trend of spread of the infection as it depends on population density, travel, and individual social aspects like cultural and life styles. Therefore, there is a need for deep learning approaches to accurately predict the COVID-19 trends in India.
	In deep learning, convolutional neural network (CNN) \cite{lecun1989generalization} is one form of deep learning architecture for processing data that has a grid like topology. It includes the time series data that can be considered as 1D grid taking samples at regular time intervals and image data considered as 2D grid of pixels. A typical end-to-end CNN network consists of different layers such as convolution, activation, max-pooling, softmax layer etc.
	
	Recurrent neural network (RNN) \cite{rumelhart1986learning} derived from the feedforward neural networks can use their interval states (memory)  to process variable length sequences of data suitable for the sequential data. Long Short-Term Memory (LSTM) has been introduced by Hochreiter and Schmidhuber \cite{hochreiter1997long} which overcomes the vanishing  and exploding gradient problem in RNN and have long dependencies that proved to be very promising for modelling of sequential data. A common LSTM unit is composed of a cell, an input gate, an output gate and a forget gate. The cell remembers values over arbitrary time intervals and the three gates regulate the flow of information into and out of the cell. For a given input sequence   $ x=(x_{1}, x_{2},\ldots x_{T}) $ from time $ t=1 $ to $ T $, LSTM calculates an output sequence $ y=y_{1},y_{2},\ldots y_{T} $, mathematically represented as \cite{hochreiter1997long}:

	\begin{equation}\label{key}
		i_{i}=\sigma(\mathbf{W}_{ix}x_{t}+\mathbf{W}_{im}m_{t-1}+\mathbf{W}_{ic}c_{t-1}+b_{i})
	\end{equation}
	\begin{equation}\label{key1}
		i_{f}=\sigma(\mathbf{W}_{fx}x_{t}+\mathbf{W}_{fm}m_{t-1}+\mathbf{W}_{fc}c_{t-1}+b_{f})
	\end{equation}
	\begin{equation}\label{key2}
		c_{t}=f_{t}\odot c_{t-1}+i_{t}\odot g(\mathbf{W}_{cx}x_{t}+\mathbf{W}_{cm}m_{t-1}+b_{c})
	\end{equation}
	\begin{equation}\label{key3}
		o_{t}=\sigma(\textbf{W}_{ox}x_{t}+\mathbf{W}_{om}m_{t-1}+\mathbf{W}_{oc}c_{t-1}+b_{o})
	\end{equation}
	\begin{equation}\label{key4}
		m_{t}=o_{t}\odot h(c_{t})
	\end{equation}
	\begin{equation}\label{key5}
		y_{t}=\phi (\mathbf{W}_{ym}m_{t}+b_{y})
	\end{equation}
	
		From Equation \ref{key} to Equation \ref{key5}, $ i,o,f $ and $ c $ represent the input gate, output gate, forget gate and cell activation vector respectively, $ m $ depicts hidden state vector also known as output vector of the LSTM unit. $ \mathbf{W} $ denotes the weight matrix, for example  $ \mathbf{W}_{ix} $ means weight matrix from input gate to input. The $ \odot $ stands for element wise multiplication, and b denotes the bias term, whereas $g$ and $h$  are used for activation functions at the input and output respectively. $ \sigma$ represents logistic sigmoid function.
	
	LSTM is a method having multiple layers which can map the input sequence to a vector having fixed dimensionality, in which the deep LSTM decodes the target sequence from the vector. This deep LSTM is essential for a recurrent neutral network model except on the input sequence. The LSTM can solve problems with long term dependencies which may be caused due to the introduction of many short term dependencies to the dataset. LSTM has the ability to learn successfully on data having a long range of temporal dependencies because of the time lag between the input and their corresponding outputs \cite{sutskever2014sequence}. LSTM can be used for predicting time series and it is beneficial for sequential data\cite{abdollahi2021modeling}. 	
	
	Deep learning models such as LSTM and CNN are well suited for understanding and predicting the dynamical trend of COVID-19 spread and have recently been used in prediction by several researchers \cite{shastri2020time, wang2020time, iqbal2021covid, bedi2021prediction, dairi2021comparative, devaraj2021forecasting, nabi2021forecasting}. Chandra et al. \cite{chandra2021deep} used the LSTM and its variants for ahead prediction of COVID-19 spread for India with split the training and testing data as static and dynamics.  LSTMs have been used for COVID-19 transmission in Canada by Chimmula \& Zhang \cite{chimmula2020time} and results show the linear transmission in the Canada .  Arora et al. \cite{arora2020prediction} performed forecasting of the COVID-19 cases for India using LSTMs variants and categorized the Indian states in different zones based on COVID-19 cases. 
	
	In this paper, we employ the vanilla LSTM, stacked LSTM, ED-LSTM, Bi-LSTM, CNN, and hybrid CNN+LSTM model to capture the dynamic trend of COVID-19 spread and predict the COVID-19 daily confirmed cases for 7, 14 and 21 days for India and its four most affected states: Maharashtra, Kerala, Karnataka, and Tamil Nadu. To demonstrate the performance of deep learning models, RMSE and MAPE errors are computed on the testing data. The flowchart of the model is represented in Figure \ref{flowchart}.
	
	The rest of the manuscript is organized as follows. Section \ref{md}, describes the deep learning model along with experimental setup and evaluation metrics. In Section \ref{rd}, we present the COVID-19 dataset and experimental results and discussions. Finally, the conclusion is made in Section \ref{con}. 

\renewcommand{\thefigure}{1}
\begin{figure*}[t] \label{flowchart}
	\centering
	\includegraphics[width=14cm, height=15cm]{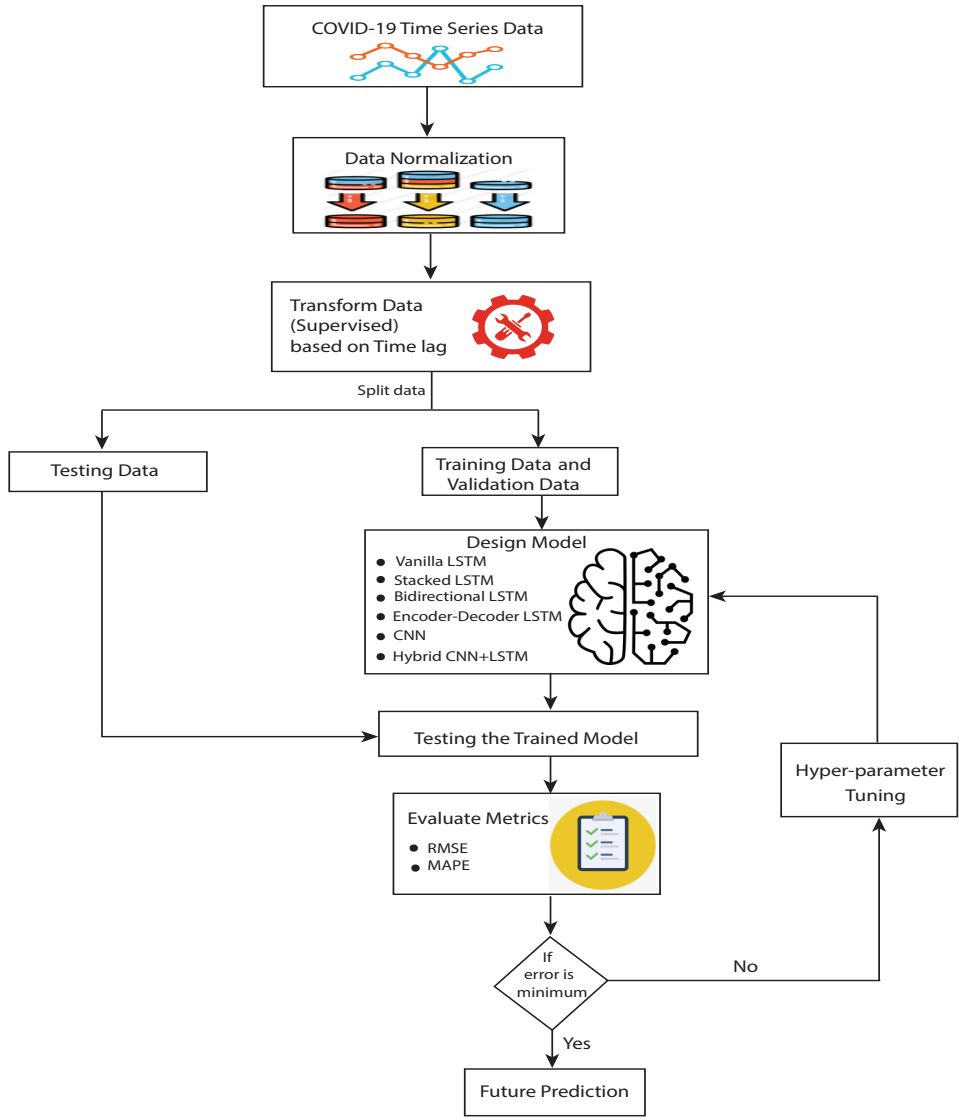}
	\caption{Flowchart describing the experimental work.}
\end{figure*}
	\section{Methods}\label{md}
	\subsection{Experimental setup}
	The COVID-19 outbreak trend is highly dynamic and depends on imposing various intervention strategies. To capture the complex trend, in this study, we proceed the following steps during the training, testing and forecasting. 
	\begin{itemize}
		\item We used early COVID-19 data up to July 10, 2021, and split the COVID-19 time series data into training and testing data by taking the last 20 days data as testing data and remaining data as training data. 
		\item To avoid the inconsistency in COVID-19 time series data, the data is normalized in the interval[0,1] using 'MinMaxScaler' Keras function. 
		\item The COVID-19 time series data is reshaped into the input shape data by taking time step (time-lag) or observation window 15 and number of features is one as for the univariate model. The observation window 15 means, we are using previous 15 days COVID-19 time series data to predict the next day, that is the 16$^{th}$ day.  In a univariate model the input contains only one feature.
		\item Further, we train and test the recurrent and convolutional neural network approaches on COVID-19 time series data and setup the model with setting hyper parameters through manual search. COVID-19 daily confirmed cases predictions are performed up to July 17, 2021(7 days), up to July 24, 2021 (14 days) and up to July 31, 2021(21 days) from July 10, 2021 using Vanilla LSTM, Stacked LSTM, ED-LSTM, Bi-LSTM, CNN, and Hybrid CNN+LSTM for India and its four most affected states Maharashtra, Kerala, Karnataka and Tamil Nadu. The experimental work is summarized in Figure 1.
	\end{itemize}
	
	RNN abd CNN approaches viz. vanilla LSTM, stacked LSTM, ED-LSTM, Bi-LSTM, CNN, and hybrid CNN+LSTM have been implemented in Python using Keras module of Tensorflow and consider the prediction by taking univariate approaches. 
	
	\subsection{Vanilla LSTM}
	A Vanilla LSTM is an LSTM model that has a single hidden layer of LSTM units. The encoder is responsible for interpreting and reading the input sequence whereas the output encoder has a fixed-length vector\cite{brownlee2018develop}. Vanilla LSTM has a property to isolate the effect due to change on the performance variant. So, when vanilla LSTM is used as a baseline it evaluates with all of its variants and allows the isolating effect for the changes made in each of the variants. The performance of vanilla LSTM is reasonably well on various data sets\cite{greff2016lstm}. This vanilla LSTM is kind of art model for different variety of machine learning programs. So, vanilla LSTM neural networks predict with accuracy making most of the long short-term memory when the cases are complicated while operating\cite{wu2018remaining}.
	\subsection{Stacked LSTM} Stacked LSTM has more than one LSTM sub-layers that are connected together using various weight parameters. On a single-layer LSTM, stacked LSTM ovelays the hidden layers of LSTM\cite{sun2020stacked}. In stacked LSTM each edge weight corresponds to weight value and the cell is the time unit. The data transformation process performed in stacked LSTM is mathematically shown below,
	\begin{equation}\label{key6}
		i^{next}=f(\sum^{M}_{n=1}(w^{next}_n * o_n + b^{next})
	\end{equation}
	Here, f is the activation function, $i^{next}$ is the input data for the next hidden layer, weight of edge connected to previous output and next layer input is defined in $w^{next}_n$, $o_m$ contains output value of one cell and $ b^{next}$ contains bias. For feature extraction the stacked LSTM proves to improve the extraction process\cite{yu2019review}. 
	\subsection{Bidirectional-LSTM}
	Bidirectional Long Short-Term memory (Bi-LSTM) is a deep learning algorithm applied for forecasting the time series data. It is adopted to learn from the framework providing better understanding from the learning context\cite{abdollahi2021modeling}. As Bi-LSTM is a multivariate time series it allows multiple time series dependent which can be designed together to predict the correlations along with the series recorded or captured variables varying simultaneously over time period\cite{said2021predicting}. Bi-LSTM is a deep learning models for the sequential prediction without much error\cite{shahid2020predictions}. It has many more features like handling temporal dependencies along with time series data distributing free learning models and flexibility in modelling non-linear features. In other words, Bi-LSTM is an enhanced version of LSTM algorithm  in which it can deal with the combination of two variants having hidden states that allows information to come from the backward layer as well as from the forward layer. The Bi-LSTM is helpful for situation that require context input. It is widely used in classification especially like text classification, sentiment classification and speed classification and recognition and load forecasting. As Bi-LSTM is a deep learning models having capacity to capture non-linearity process and being flexible in modelling time-dependent data; so now-a-days Bi-LSTM have been using for real-time forecasts of the daily events\cite{zeroual2020deep}.
	\subsection{EncoderDecoder-LSTM}
	ED-LSTM (Encoder Decoder) is a network of sequence-to-sequence model for mapping a fixed-length input to a fixed-length output. It handles variable length input and output first by encoding the input sequence, then decoded from the representation. This method can compute a sequence having hidden states. In  ED-LSTM, the encoder and decoder improved the continuity of learning input and output sequences. It experiences reuse for reading input sequence and writing output sequence many times sequentially. And the times of reuse skill depend on the length of the input and output sequences. ED-LSTM model is so consistent and its outputs are stable, reliable and accurate. It can even effectively mimic the long-term dependence between variables\cite{kao2020exploring}. The advantage of ED-LSTM is that the network of models can be constructed from the model definition which consists of a list of input and outputs. So, the models can be automatically trained from the provided dataset. This advantage of ED-LSTM help to reduce the model construction and training cost\cite{ellis2020encoder}.

	\subsection{Convolution Neural Network(CNN)}
	CNN is one of the algorithms in Deep learning that automatically captures and identifies the important features without the need of human supervision\cite{gu2018recent}. Local connections and shared weights employed in the CNN are useful in extracting features from 2-D input signals such as image signals. Basically, CNN has three kinds of layers: convolution layer, pooling layer and fully connected layer. Convolution layer is primarily associated with the identification of features from raw data. This is achieved by applying filters having predefined size followed by convolution operation.  Pooling layer applies a pooling operation that reduces the dimension of feature maps while retaining the important features\cite{albawi2017understanding}. Some of the pooling methods are max pooling and average pooling. The fully connected layer or the dense layer generates forecasting after features extracting process. The final fully connected layers have flattened features arising after applying convolution and pooling operations\cite{alzubaidi2021review,zhou2016recurrent}.  
\\
(I) The convolutional layer in CNN architecture consists of multiple convolutional filters. These filters are also known as kernels. Convolution operation is performed between the raw data that is in the form of a matrix and these kernels that generate an output feature map. The numbers present in the kernel is the kernel weight of the kernel. The initial values of the kernel are random in nature, during the training process the kernel values are adjusted to help in extracting important features from the data. In convolutional operation the CNN input format description is present. In convolution operation let’s say in 10$*$10 grey-scale image a randomly initialized kernel slides vertically and horizontally and the dot product between them is computed.  In 1D-CNN the kernel function moves in one direction only. Similarly, in 2D-CNN and 3D-CNN the kernel function moves in two and three directions respectively. The computed values are multiplied to create a single scalar value. 
	The data processed by the kernel of CNN sometimes may require padding. It is a process of extracting border information from the input data. Padding refers to the adding layers of extra pixels (zeros) to the input data that helps to preserve information present on the borders\cite{albawi2017understanding}.
\\
(II) Pooling layer: The feature maps generated from the convolutional operations are sub-sampled in the pooling layer. This reduces the large size feature maps to generate smaller feature maps. The pooling layer reduces the dimension of the feature map resulting in reduction in the number of parameters to learn. It also reduces the computation that needs to be performed. There are various types of pooling such as average pooling, max pooling, min pooling, global average pooling (GAP) etc. It may be possible sometimes that the performance output of CNN model decreases because of the pooling layer as it focuses primarily on ascertaining the correct location of a feature rather than focusing on particular features available in the data\cite{gu2018recent,alzubaidi2021review,zhou2016recurrent}.
\\
(III). Activation function (Transfer function):  In a neural network based on the weighted sum of the neuronal input activation function transforms it into output form. It performs mapping of the input to the output depending upon the neuronal input so as to fire a particular neuron or not.  Activation functions can be linear or non-linear functions. Some of the activation function used in CNN are described below:
	(a)Rectilinear Unit (ReLU): The ReLU function converts the input to a piecewise linear function to a positive output otherwise it will output zero. It is one of the common activation functions in most of the neural networks. One of the advantages of using ReLU over other activation functions is that it has lower computational load\cite{albawi2017understanding}. Mathematically it is represented as below,
	\begin{equation}\label{key9}
		f(x)_{ReLU}=max\{0,x\}
	\end{equation}
	(b)Sigmoid: In this the input are real numbers and the output is constrained to be in between zero and one. It is S-Shaped function and is mathematically represented as shown below,
	\begin{equation}\label{key10}
		f(x)_{sigmoid}=\frac{1}{1+e^{-x}}
	\end{equation}
	(c)Tanh: In Tanh activation function the input is real numbers and output is in between -1 and 1. It is described mathematically as shown below,
	\begin{equation}\label{key11}
		f(x)_{tanh}=\frac{e^x - e^{-x}}{e^x + e^{-x}}
	\end{equation}
\\
(IV). Fully Connected layer: In this layer each neuron is fully connected to other neurons of the other layer, hence the name Fully Connected (FC) layer. It is located at the end of the CNN architecture and it forms the last few layers in the network. The final pooling layer that is flattened is the input to the FC layer. Flattening is a process in which a matrix is unrolled at its values to form a vector\cite{albawi2017understanding}.
\\
(V). Loss function: Loss functions are used in the output layer to compute the predicted error created during training samples in CNN. This error is the difference between the actual output and the predicted values. Some of the loss functions used in neural network are Mean Squared Error (MSE), Cross-Entropy or Softmax loss function, Euclidean loss function and Hinge loss function\cite{albawi2017understanding}.

	\subsection{Hybrid CNN+LSTM}
	Hybrid CNN+LSTM deep learning architecture combines the benefits of both the LSTM and CNN. The LSTM in this hybrid model learns the temporal dependencies that are present in the input data. The CNN is integrated such that it can process high dimensional data. The components of LSTM are input gate, forget gate, output gate, memory cell, candidate memory cell and hidden state\cite{li2020hybrid}.   The 1-D CNN finds the important features from temporal feature space using non-linear transformation generated by LSTM.  The convolution layers are wrapped with a time-distributed layer in the model and it is ensured that data is transformed appropriately.  The layers used in the model are two convolutional layers, max-pooling layer, flatten layer, time-distributed layer, followed by LSTM layers\cite{li2020hybrid,liu2018hybrid}.
	
	\subsection{Evaluation Metrics}
	To demonstrate the relative performance of various deep learning models, the root mean square error (RMSE) and mean absolute percentage error (MAPE) have been computed, which is mathematically defined as:
	\begin{equation}\label{key12}
		RMSE=\sqrt{1/n{\sum^{n}_{i=1}(y_i - \bar{y})^2}}
	\end{equation}
	\begin{equation}\label{key13}
		MAPE=\frac{100}{n} \left|\frac{(y_i - \bar{y})^2}{y_i}\right|
	\end{equation}
	here $y_i$ denote the actual confirmed cases, $\bar{y_i}$ is the predicted daily confirmed cases using the deep learning model, and n is the total number of observation under the study. The small value of RMSE and MAPE represents the better performance of that model. In this study, RMSE and MAPE are computed on the test data where the actual and predicted values of various other models are available. Throughout all predictions of 7, 14, and 21 days, we also computed the confidence interval \cite{gupta1994fundamental} at 95\% for the predicted new confirmed COVID-19 cases counts per day. The confidence interval gives a range of values for new cases and it gives the probability with which an estimated interval will contain the true value of the confirmed cases.
	\section{Results and Discussions}\label{rd}
	For the analysis and forecasting of the daily confirmed COVID-19 cases for training and testing the RNN and CNN models are considered. In our study we used Vanilla LSTM, Stacked LSTM, ED-LSTM, Bi-LSTM, CNN, and Hybrid CNN+LSTM to build a map that captures the complex trend in the given sequence of COVID-19 time series data and performs forecasting using these maps. The details are discussed in the following subsections: 
	\subsection{COVID-19 data and preprocessing}
	In this study, daily new COVID-19 cases have been predicted for 7 days, 14 days and 21 days for the whole country (India) and four of its most affected states (Maharashtra, Kerala, Karnataka, and Tamil Nadu) using deep learning approaches. Previous COVID-19 time series data is accessed from COVID-19India.org during January 30, 2020 to July 10, 2021, where numbers of daily confirmed, recovered and deceased cases are publicly available online at \url{https://api.covid19india.org/documentation/csv/}. We use data up to July 10, 2021 as illustrated in Fig. 2(a-e) to train and test the recurrent and convolutional  neural network models. The trends of COVID-19 time series data is highly inconsistent in nature and it may be due to the rate of individual infections, number of reporting of the cases, individual behaviour, effect of lockdown, and non- pharmaceuticals measures. India and its states witnessed two waves and new cases count per day during peak of the second wave were much more than the first wave as depicted in Fig.2 . Due to higher consistency in per day count, these data are normalized in the interval of [0,1] using ‘MinMaxScaler’ of the keras function in the preprocessing step before applying the deep learning models. 
	
	The ‘MinMaxScaler’ function normalizes the given time series data $(x)$ using the formula $x_{normal}=(x-x_{min})/(x_{max}-x_{min})$, where $x_{max}$ and $x_{min}$ represents the maximum and minimum value of data $(x)$. After the forecasting of the confirmed cases count per day that lies in the interval $[0,1]$ it is again re-transformed into the corresponding actual number by applying reverse operation using $‘inverse\_transform’$ keras function.
	\renewcommand{\thefigure}{2(a)}
	\begin{figure}
		\centering
		\includegraphics[width=90mm,height=45mm]{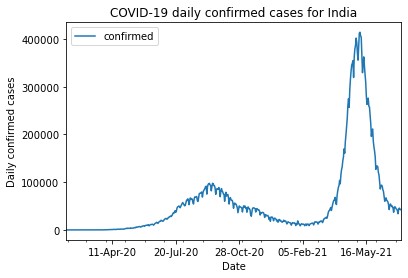}
		\vspace*{-9mm}
		\caption{Daily confirmed COVID-19 time series data for India from Jan 30, 2020 to Jul 10, 2021}
	\end{figure}
	\renewcommand{\thefigure}{2(b)}
	\begin{figure}
		\centering
		\includegraphics[width=90mm,height=45mm]{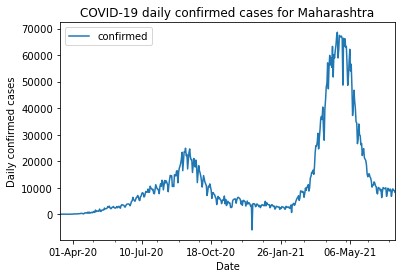}
		\vspace*{-9mm}
		\caption{Daily confirmed COVID-19 time series data for Maharashtra from Mar 14, 2020 to Jul 10, 2021}
	\end{figure}
	
	\renewcommand{\thefigure}{2(c)}
	\begin{figure}
		\centering
		\includegraphics[width=90mm,height=50mm]{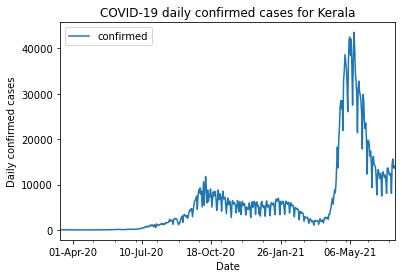}
		\vspace*{-9mm}
		\caption{Daily confirmed COVID-19 time series data for Kerala from Mar 14, 2020 to Jul 10, 2021}
	\end{figure}
	\renewcommand{\thefigure}{2(d)}
	\begin{figure}
		\centering
		\includegraphics[width=90mm,height=45mm]{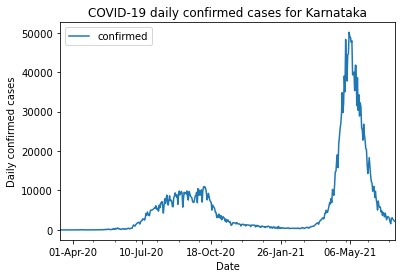}
		\vspace*{-9mm}
		\caption{Daily confirmed COVID-19 time series data for Karnataka from Mar 14, 2020 to Jul 10, 2021}
	\end{figure}
	\renewcommand{\thefigure}{2(e)}
	\begin{figure}
		\centering
		\includegraphics[width=90mm,height=50mm]{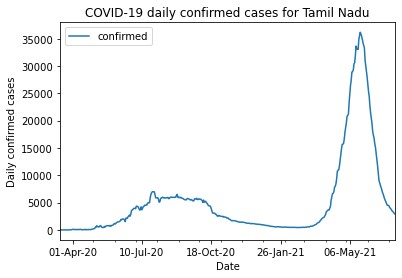}
		\vspace*{-9mm}
		\caption{Daily confirmed COVID-19 time series data for Tamil Nadu from Mar 14, 2020 to Jul 10, 2021}
	\end{figure}
	\subsection{Hyper parameter tuning}
	The hyper parameters in vanilla LSTM, stacked LSTM, ED-LSTM, Bi-LSTM, CNN, and hybrid CNN+LSTM models are summarized in Table \ref{hpsl} and Table \ref{hyper}. To avoid the over-fitting, we regularize the model on the training data using $ L1 $\ regularizer (bias/kernel) with different settings along with Dropout as shown in Table \ref{hpsl} and Table \ref{hyper}. Around 20\% to 40\% neurons are dropped through the Dropout layers. In CNN and hybrid CNN+LSTM, we use the Conv1D layer along with the kernel size\ 2, depicted in Table \ref{hyper}. Throughout the entire experiment 'ReLu' activation function, 'adamax' optimizer and 'MSE' loss function is considered in our study. As tuning the training epochs, we setup the 'EarlyStopping' callback with number of epochs 1000, batch size 64 along with patience=250. This setup checks the performance of the respective model on train and validation datasets and stops the training if it looks like that if the model is starting to over learn or over fit. The learning algorithm is stochastic in nature therefore the results may be varying in nature\cite{brownlee2018better}. To address this issue, we have run each deep learning model up to 10 times and saved the better model and noted their corresponding performance results in our experiment.
\begin{table}[htbp]
	\centering
	\caption{Hyper-parameter and structure of vanilla LSTM, stacked LSTM, ED-LSTM, BiLSTM models}
	\scalebox{0.6}{
		\begin{tabular}{rlccrlcc}
			\toprule
			\toprule
			\multicolumn{1}{r}{\multirow{3}[2]{*}{Models}} & \multirow{3}[2]{*}{Layers} & \multicolumn{1}{l}{Number } & \multicolumn{1}{l}{Bias} & \multicolumn{1}{r}{\multirow{3}[2]{*}{Models}} & \multirow{3}[2]{*}{Layers} & \multicolumn{1}{l}{Number } & \multicolumn{1}{l}{Bias} \\
			&     & \multicolumn{1}{l}{of units} & \multicolumn{1}{l}{regularizer } &     &     & \multicolumn{1}{l}{of units} & \multicolumn{1}{l}{regularizer } \\
			&     &     & \multicolumn{1}{l}{L1} &     &     &     & \multicolumn{1}{l}{L1} \\
			\midrule
			\midrule
			\multicolumn{1}{l}{Vanilla LSTM} & LSTM & 200 & 0.02 & \multicolumn{1}{l}{BiLSTM} & Bidirectional & 250 & 0.4 \\
			& Dropout & 0.2 & -   &     & Dropout & 0.2 & - \\
			& Dense & 1   & -   &     & Dense & 1   & - \\
			\multicolumn{1}{l}{Stacked LSTM} & LSTM & 130 & 0.04 & \multicolumn{1}{l}{ED\_LSTM} & LSTM & 125 & 0.02 \\
			& Dropout & 0.4 & -   &     & Dropout & 0.2 & - \\
			& LSTM & 100 & 0.04 &     & Repeat Vector & 4   & - \\
			& Dropout & 0.4 & -   &     & LSTM & 75  & 0.02 \\
			& LSTM & 75  & 0.04 &     & Dropout & 0.2 & - \\
			& Dense & 1   & -   &     & Dense & 1   & - \\
			\bottomrule
	\end{tabular}}%
	\label{hpsl}%
\end{table}%
\begin{table}[htbp]
	\centering
	\caption{Hyper-parameter and structure of CNN, and CNN+LSTM models}
	\scalebox{0.73}{
		\begin{tabular}{rlcccc}
			\toprule
			\toprule
			\multicolumn{1}{r}{\multirow{3}[2]{*}{Models}} & \multirow{3}[2]{*}{Layers} & \multicolumn{1}{l}{ Number } & \multicolumn{1}{l}{kernel } & \multicolumn{1}{l}{Bias} & \multicolumn{1}{l}{Kernel} \\
			&     & \multicolumn{1}{l}{of filters/} & \multicolumn{1}{l}{size} & \multicolumn{1}{l}{regularizer } & \multicolumn{1}{l}{regularizer} \\
			&     & \multicolumn{1}{l}{units} &     & \multicolumn{1}{l}{L1} & \multicolumn{1}{l}{ L1} \\
			\midrule
			\midrule
			\multicolumn{1}{l}{CNN} & Conv1D & 100 & 2   & 0.4 & 0.002 \\
			& Conv1D & 75  & 2   & 0.4 & 0.002 \\
			& MaxPooling1D  & -   & 2   & -   & - \\
			& Flatten & -   & -   & -   & - \\
			& Dense & 64  & -   & -   & - \\
			& Dense & 1   & -   & -   & - \\
			\multicolumn{1}{l}{CNN+LSTM} & Conv1D & 100 & 2   & -   & 0.002 \\
			& MaxPooling1D  & -   & 2   & -   & - \\
			& Flatten & -   & -   & -   & - \\
			& LSTM & 64  & -   & 0.5 & - \\
			& Dropout & 0.3 & -   & -   & - \\
			& Dense & 1   & -   & -   & - \\
			\bottomrule
	\end{tabular}}%
	\label{hyper}%
\end{table}%

\renewcommand{\thefigure}{3(a)}
\begin{figure}
	\centering
	\includegraphics[width=90mm,height=70mm]{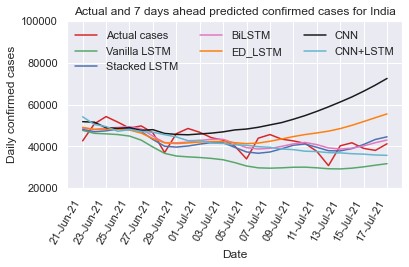}
	\vspace*{-9mm}
	\caption{Predicted and actual cases for India ahead of 7 days}
\end{figure}
\renewcommand{\thefigure}{3(b)}
\begin{figure}
	\centering
	\includegraphics[width=90mm,height=70mm]{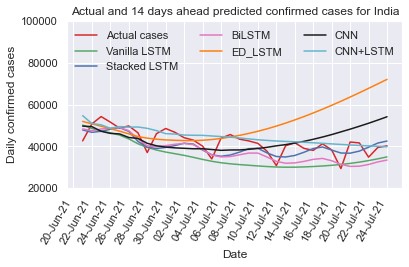}
	\vspace*{-9mm}
	\caption{Predicted and actual cases for India ahead of 14 days}
\end{figure}
\renewcommand{\thefigure}{3(c)}
\begin{figure}
	\centering
	\includegraphics[width=90mm,height=70mm]{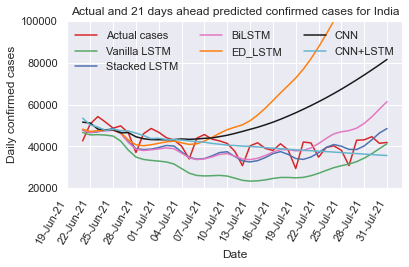}
	\vspace*{-9mm}
	\caption{Predicted and actual cases for India ahead of 21 days}
\end{figure}
\subsection{Prediction performance}
In this section, we discuss the prediction performance of deep learning models for India and four it states: Maharashtra, Kerala, Karnataka, and Tamil Nadu, individually in the following subsections:
\subsubsection{India}
India is the second most populous country in the world, it may lead to higher threats because of the spread of COVID-19. The daily confirmed cases in India from Jan 30, 2020 to July 10, 2021 are depicted in Fig. 2(a). It is observed that the new confirmed cases per day are highly inconsistent. India witnessed two waves, in second waves around 400,000 new cases were reported per day. To address this issues, we train and test the vanilla LSTM, stacked LSTM, ED-LSTM, Bi-LSTM, CNN, and CNN+LSTM models on the India normalized time series data to capture the real trend with setting the hyper parameter, shown in Table \ref{hpsl} and Table \ref{hyper}, with manual tuning of hyper parameters. The predicted new cases of COVID-19 for 7, 14, and 21 days are calculated from July 10, 2021 using various recurrent and CNN model and determine the corresponding performance metrics: RMSE and MAPE as presented in Table 3. RMSE and MAPE are computed for the actual and predicted daily confirmed case from June 21, 2021 to July 10, 2021 on the test data. From Table \ref{rmse}, it can be seen that the RMSE and MAPE (7.57\% - 11.36\%) are comparatively smaller for the stacked LSTM and hybrid CNN+LSTM. In some cases RMSE and MAPE (7.36\%-12.96\%) is less for Bi-LSTM and ED-LSTM on the test data but the predicted new cases per day is far from the actual cases (Fig. 3). Bi-LSTM and ED-LSTM models have the over-fitting problem. The predicted and actual (red color) cases for India for 7 days (up to July 17, 2021), 14 days (up to July 24, 2021) and 21 days (up to July 31, 2021) are shown in Figs. 3(a)-3(c). It can be observed that the stacked LSTM and hybrid CNN+LSTM provide better prediction as forecasting in count cases is close to actual count per day. The predicted new cases for 7, 14 and 21 days with various models along with 95\% level confidence intervals are shown in Table 4. In our study, for India, we found that stacked LSTM and hybrid CNN+LSTM performed best in terms of prediction consistency among all six deep learning models.
\begin{table*}[t]
	\centering
	\caption{RMSE and MAPE with various model from June 21, 2021 to July 10, 2021}
	\scalebox{0.8}{
		\begin{tabular}{llcccccc}
			\toprule
			\toprule
			\multirow{2}[2]{*}{} & \multirow{2}[2]{*}{} & \multicolumn{2}{l}{Next 7 days } & \multicolumn{2}{l}{Next 14 days } & \multicolumn{2}{l}{Next 21 days } \\
			&     & \multicolumn{2}{l}{prediction} & \multicolumn{2}{l}{prediction} & \multicolumn{2}{l}{prediction} \\
			\midrule
			Country/States & Models & \multicolumn{1}{r}{RMSE} & \multicolumn{1}{r}{MAPE} & \multicolumn{1}{r}{RMSE} & \multicolumn{1}{r}{MAPE} & \multicolumn{1}{r}{RMSE} & \multicolumn{1}{r}{MAPE} \\
			\midrule
			\midrule
			\multirow{6}[1]{*}{India} & Vanilla LSTM & 9746.6 & 19.57 & 8454.28 & 17.01 & 11912.86 & 24.09 \\
			& Stacked LSTM & 4687.34 & 8.68 & 5273.4 & 9.99 & 6023.1 & 11.36 \\
			& BiLSTM & 3889.28 & 7.34 & 5349.35 & 10.3 & 6369 & 12.29 \\
			& ED\_LSTM & 4067.74 & 7.95 & 4385.09 & 8.1 & 4431.91 & 8.75 \\
			& CNN & 6729.1 & 12.61 & 5320.62 & 10.89 & 4304.57 & 7.6 \\
			& CNN+LSTM & 4831.59 & 9.07 & 4916.03 & 7.86 & 4431.45 & 7.57 \\
			\multirow{6}[0]{*}{Maharashtra} & Vanilla LSTM & 1455.74 & 14.63 & 2594.56 & 25.73 & 1075.67 & 10.69 \\
			& Stacked LSTM & 1505.12 & 15.55 & 1046.71 & 10.21 & 1125.31 & 10.8 \\
			& BiLSTM & 1038.16 & 9.95 & 1339.5 & 14.08 & 1041.19 & 10.24 \\
			& ED\_LSTM & 1889.96 & 19.36 & 1288.35 & 13.8 & 1363.88 & 14.56 \\
			& CNN & 1871.02 & 19.54 & 1140.99 & 11.5 & 1232.49 & 12.65 \\
			& CNN+LSTM & 1413.44 & 13.61 & 1231.39 & 11.83 & 1328.62 & 12.76 \\
			\multirow{6}[0]{*}{Kerala} & Vanilla LSTM & 3758.49 & 22.17 & 2219.32 & 13.78 & 1470.86 & 9.55 \\
			& Stacked LSTM & 3747.23 & 23.48 & 2740.37 & 19.24 & 1801.96 & 13.2 \\
			& BiLSTM & 2468.36 & 18.21 & 3015.41 & 21.99 & 1778.38 & 13.56 \\
			& ED\_LSTM & 2436.79 & 17.05 & 2700.32 & 19.71 & 2304.13 & 16.57 \\
			& CNN & 1836.93 & 13.6 & 1839.97 & 13.58 & 2436.62 & 17.26 \\
			& CNN+LSTM & 2641.06 & 19.17 & 1950.78 & 13.54 & 4276.85 & 28.01 \\
			\multirow{6}[0]{*}{Karnataka} & Vanilla LSTM & 948.58 & 27.13 & 461.05 & 14.14 & 460.7 & 13.72 \\
			& Stacked LSTM & 616.77 & 19.12 & 482.77 & 13.43 & 876.11 & 25.35 \\
			& BiLSTM & 702.86 & 21.95 & 632.44 & 19.64 & 549.77 & 17.01 \\
			& ED\_LSTM & 495.17 & 15.81 & 497.36 & 15.42 & 720.96 & 21.64 \\
			& CNN & 539.86 & 17.15 & 767.04 & 22.86 & 767.9 & 22.64 \\
			& CNN+LSTM & 659.5 & 21.37 & 513.75 & 15.96 & 590.59 & 18.37 \\
			\multirow{6}[1]{*}{Tamil Nadu} & Vanilla LSTM & 1096.36 & 25.25 & 1437.03 & 33.79 & 1571.95 & 36.87 \\
			& Stacked LSTM & 1527.06 & 35.82 & 671.79 & 14.99 & 1253.73 & 29.34 \\
			& BiLSTM & 1716.46 & 40.13 & 393.62 & 8   & 779.47 & 17.67 \\
			& ED\_LSTM & 700.86 & 14.66 & 768.71 & 15.97 & 826.78 & 17.26 \\
			& CNN & 1516.43 & 35.44 & 1040.41 & 24.12 & 1134.7 & 26.52 \\
			& CNN+LSTM & 594.12 & 11.8 & 823.92 & 15.37 & 1419.17 & 31.53 \\
			\bottomrule
	\end{tabular}}%
	\label{rmse}%
\end{table*}%

\subsubsection{Maharashtra}
Maharashtra was one of the worst-affected states in India during the second wave with COVID-19. The new cases count per day is depicted in Fig. 2(b), which shows that the number of daily cases might count nearly 70,000 in the second waves and outbreak scenario being highly dynamic. To capture the dynamic trend of data, we train and test the vanilla LSTM, stacked LSTM, ED-LSTM, BiLSTM, CNN, and hybrid CNN+LSTM model on Maharashtra time series data with setting the hyper parameters, illustrated in Table 1 and Table 2, and computed RMSE and MAPE on test data, presented in Table 3. Further forecasting of confirmed new cases per day for 7 days (up to July 17, 2021), 14 days (up to July 24, 2021) and 21 days (up to July 31, 2021) from July 10, 2021 are shown in Table 4. Fig. 4(a)-(c) illustrates the predicted and actual cases using deep learning models. In 7 days prediction, the stacked LSTM (MAPE=15.55\%) and Bi-LSTM (MAPE=9.95\%) forecasts value close to the actual values whereas in 14 days prediction the Bi-LSTM and ED-LSTM forecasts cases close to actual cases. Table 4 shows 95\% confidence interval for the predicted confirmed cases per day up to July 31, 2021.
\renewcommand{\thefigure}{4(a)}
\begin{figure}
	\centering
	\includegraphics[width=90mm,height=70mm]{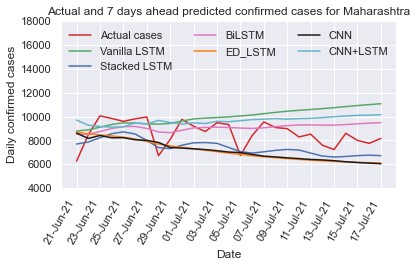}
	\vspace*{-9mm}
	\caption{Predicted and actual cases for Maharashtra ahead of 7 days}
\end{figure}
\renewcommand{\thefigure}{4(b)}
\begin{figure}
	\centering
	\includegraphics[width=90mm,height=70mm]{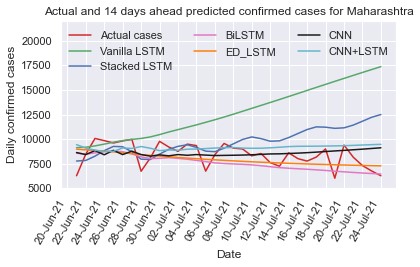}
	\vspace*{-9mm}
	\caption{Predicted and actual cases for Maharashtra ahead of 14 days}
\end{figure}
\renewcommand{\thefigure}{4(c)}
\begin{figure}
	\centering
	\includegraphics[width=90mm,height=70mm]{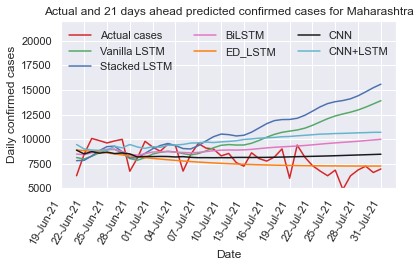}
	\vspace*{-9mm}
	\caption{Predicted and actual cases for Maharashtra ahead of 21 days}
\end{figure}

\begin{table*} [t]
	\centering
	\caption{Prediction of new confirmed case ahead of 7, 14 and 21 days with various models}
	\scalebox{0.9}{
		\begin{tabular}{llllllll}
			\toprule
			\toprule
			&     & \multicolumn{2}{l}{Next 7 days prediction} & \multicolumn{2}{l}{Next 14 days prediction} & \multicolumn{2}{l}{Next 21 days prediction} \\
			\cmidrule{1-7}    Country/ & \multirow{3}[2]{*}{Models} & predicted  & 95\% confidence & predicted  & 95\% confidence & predicted & 95\% confidence  \\
			States &     & on  & interval & on  & interval &  on  & interval \\
			&     & \multicolumn{1}{r}{17-07-2021} &     & \multicolumn{1}{r}{24-07-2021} &     & \multicolumn{1}{r}{31-07-2021} &  \\
			\midrule
			\midrule
			\multirow{6}[1]{*}{India} & Vanilla LSTM & 31759 & [29437, 30832] & 35003 & [30647, 32344] & 41310 & [26377, 30991] \\
			& Stacked LSTM & 44617 & [38520, 42429] & 42595 & [36814, 39217] & 48593 & [36012, 39710] \\
			& BiLSTM & 43071 & [39296, 41623] & 33485 & [31856, 33257] & 61464 & [39453, 46424] \\
			& ED\_LSTM & 55588 & [48027, 53046] & 72134 & [55317, 63396] & 156472 & [75427, 104118] \\
			& CNN & 72572 & [59971, 68365] & 54217 & [43309, 48319] & 81682 & [56401, 65913] \\
			& CNN+LSTM & 35746 & [36094, 37051] & 39876 & [40839, 41877] & 35664 & [37370, 38641] \\
			\multirow{6}[0]{*}{Maharashtra} & Vanilla LSTM & 11080 & [10708, 10971] & 17372 & [14738, 16075] & 13909 & [10731, 11978] \\
			& Stacked LSTM & 6729 & [6812, 6656] & 12500 & [10553, 11445] & 15587 & [11992, 13394] \\
			& BiLSTM & 9497 & [9303, 9429] & 6446 & [6696, 6974] & 9979 & [9229, 9532] \\
			& ED\_LSTM & 6109 & [6149, 6279] & 7278 & [7377, 7502] & 7254 & [7291, 7356] \\
			& CNN & 6057 & [6131, 6332] & 9113 & [8616, 8847] & 8459 & [8199, 8295] \\
			& CNN+LSTM & 10164 & [9945, 10115] & 9464 & [9225, 9343] & 10693 & [10213, 10461] \\
			\multirow{6}[0]{*}{Kerala} & Vanilla LSTM & 5296 & [5390, 6225] & 17549 & [11776, 14275] & 38564 & [23776, 29651] \\
			& Stacked LSTM & 6866 & [6736, 7182] & 34136 & [23184, 28533] & 25395 & [17505, 20351] \\
			& BiLSTM & 21663 & [17734, 20477] & 30926 & [21760, 25726] & 15940 & [14720, 15209] \\
			& ED\_LSTM & 10229 & [10010, 10313] & 25344 & [20536, 22902] & 11223 & [11206, 11425] \\
			& CNN & 18715 & [16398, 17947] & 23336 & [18276, 20758] & 12219 & [11081, 11496] \\
			& CNN+LSTM & 9224 & [9366, 9829] & 17342 & [14624, 15881] & 4495 & [4828, 5390] \\
			\multirow{6}[0]{*}{Karnataka} & Vanilla LSTM & 1156 & [1128, 1154] & 2537 & [2246, 2358] & 2627 & [2423, 2490] \\
			& Stacked LSTM & 3688 & [3275, 3541] & 1581 & [1732, 1921] & 1107 & [1014, 1073] \\
			& BiLSTM & 2525 & [2575, 2719] & 1611 & [1563, 1607] & 4128 & [2505, 3068] \\
			& ED\_LSTM & 3725 & [3095, 3509] & 5910 & [3130, 4307] & 1730 & [1315, 1438] \\
			& CNN & 3966 & [3130, 3685] & 1959 & [1552, 1712] & 4246 & [2200, 2916] \\
			& CNN+LSTM & 747 & [815, 1120] & 1480 & [1607, 1773] & 1762 & [2131, 2402] \\
			\multirow{6}[1]{*}{Tamil Nadu} & Vanilla LSTM & 2093 & [1096, 2048] & 980 & [996, 1115] & 2540 & [1443, 1810] \\
			& Stacked LSTM & 1036 & [1055, 1196] & 1923 & [2042, 2209] & 551 & [814, 1063] \\
			& BiLSTM & 832 & [895, 1067] & 3247 & [3358, 3493] & 977 & [1326, 1650] \\
			& ED\_LSTM & 6263 & [4765, 5754] & 7711 & [5521, 6578] & 13661 & [7504, 9848] \\
			& CNN & 1078 & [1034, 1086] & 1698 & [1677, 1749] & 537 & [683, 908] \\
			& CNN+LSTM & 3923 & [ 3618, 4018] & 2317 & [2485, 2664] & 278 & [826, 1232] \\
			\bottomrule
	\end{tabular}}%
	\label{pred}%
\end{table*}%

\subsubsection{Kerala}
We train and test the different recurrent and convolution neural network models: vanilla LSTM, stacked LSTM, ED-LSTM, Bi-LSTM, CNN, and CNN+LSTM models on Kerala COVID-19 early data from Mar 14, 2020 to Jul 10, 2021 (Fig. 2(c)) with setting the hyper parameters (Table \ref{hpsl} and Table \ref{hyper}) to capture the trend of daily confirmed cases and computed RMSE and MAPE (Table \ref{rmse}) on test data (last 20 days data). The RMSE and MAPE (=9.55\%) values for vanilla LSTM is smallest on test data among six models. Using different learning models the prediction of 7 days (up to July 17, 2021), 14 days (up to July 24, 2021) and 21 days (up to July 31, 2021) has been done as shown in Table \ref{pred} and their comparison is illustrated in Figs. 5(a)-(c). Due to the highly dynamic trend (zigzag) of the Kerala time series data it is difficult to capture its trend. In 7 and 14 days prediction, CNN+LSTM forecasts the confirmed cases per day close to the actual cases counts per day and in 21 days prediction the stacked LSTM forecasting value is close to actual values. 
\renewcommand{\thefigure}{5(a)}
\begin{figure}
	\centering
	\includegraphics[width=90mm,height=55mm]{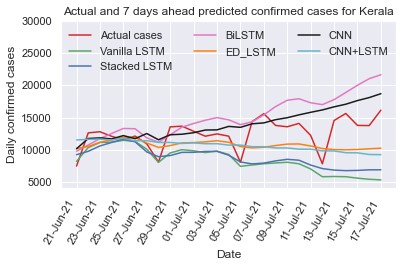}
	\vspace*{-9mm}
	\caption{Predicted and actual cases for Kerala for 7 days }
\end{figure}
\renewcommand{\thefigure}{5(b)}
\begin{figure}
	\centering
	\includegraphics[width=90mm,height=55mm]{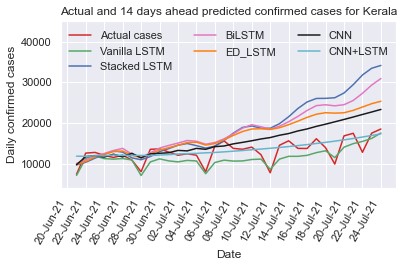}
	\vspace*{-9mm}
	\caption{Predicted and actual cases for Kerala for 14 days }
\end{figure}
\renewcommand{\thefigure}{5(c)}
\begin{figure}
	\centering
	\includegraphics[width=90mm,height=55mm]{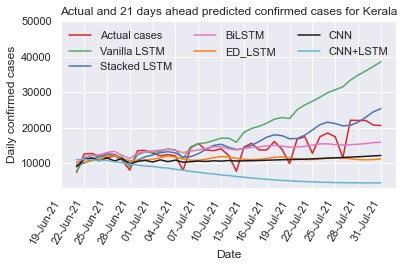}
	\vspace*{-9mm}
	\caption{Predicted and actual cases for Kerala for 21 days }
\end{figure}
\subsubsection{Karnataka}
The time series data of Karnataka depicted in Fig. 2(d) shows the dynamic trend of data during the first and the second wave. To address these issues and capture the trend of new cases count per day, vanilla LSTM, stacked LSTM, ED-LSTM, Bi-LSTM, CNN, and hybrid CNN+LSTM models are trained and tested on Karnataka data with the hyper parameters shown in Table \ref{hpsl} and Table \ref{hyper}.  Further prediction is performed for 7 days (up to July 17, 2021), 14 days (up to July 24, 2021) and 21days (up to July 31, 2021) as displayed in Table \ref{pred}. The comparisons between the predicted and actual case by different models are illustrated in Figs. 6(a)-(c). In 14 days prediction stacked LSTM gives less MAPE (=13.43\%) error among other models and also predicted new cases per day close to the actual cases, whereas in 7 days prediction the hybrid CNN+LSTM provides predicted cases per day close to actual cases. The ED-LSTM performance is better in 21 days prediction but in 14 days prediction the predicted cases are far from the actual cases that may be because of over fitting.
\renewcommand{\thefigure}{6(a)}
\begin{figure}
	\centering
	\includegraphics[width=90mm,height=55mm]{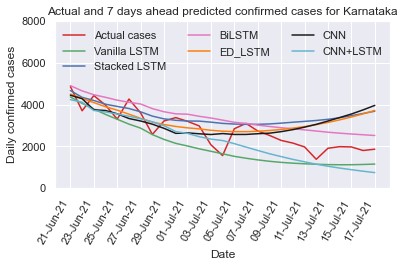}
	\vspace*{-9mm}
	\caption{Predicted and actual cases for Karnataka for 7 days }
\end{figure}
\renewcommand{\thefigure}{6(b)}
\begin{figure}
	\centering
	\includegraphics[width=90mm,height=55mm]{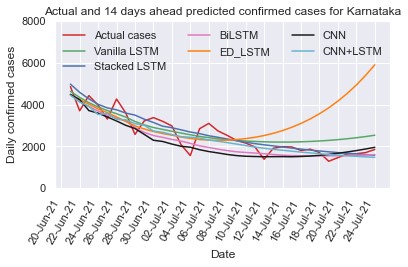}
	\vspace*{-9mm}
	\caption{Predicted and actual cases for Karnataka for 14 days }
\end{figure}
\renewcommand{\thefigure}{6(c)}
\begin{figure}
	\centering
	\includegraphics[width=90mm,height=70mm]{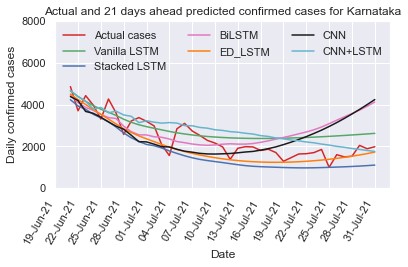}
	\vspace*{-9mm}
	\caption{Predicted and actual cases for Karnataka for 21 days }
\end{figure}
\renewcommand{\thefigure}{7(a)}
\begin{figure}
	\centering
	\includegraphics[width=90mm,height=70mm]{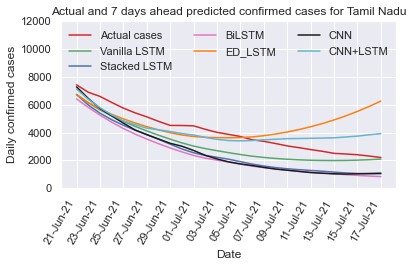}
	\vspace*{-9mm}
	\caption{Predicted and actual cases for Tamil Nadu for 7 days}
\end{figure}

\subsubsection{Tamil Nadu}
The new cases count per day in Tamil Nadu is depicted in Fig. 2(e) which shows that the number of daily cases might count nearly 35,000 in the second wave and outbreak scenario being inconsistent in nature. Further forecasting of new confirmed cases per day for 7 days (up to July 17, 2021), 14 days (up to July 24, 2021) and 21 days (up to July 31, 2021) from July 10, 2021 is shown in Table \ref{pred}. The comparison of predicted and actual cases per day for 7, 14, and 21 days using deep learning models are illustrated in Fig. 7(a)-(c). All models except ED-LSTM are able to capture the declining cases in Tamil Nadu. In 14 and 21 days prediction, forecasting of the case counts per day by vanilla LSTM, stacked LSTM, Bi-LSTM, CNN, and hybrid CNN+LSTM models are close to actual cases (Fig. 7(a)-(c)).

\renewcommand{\thefigure}{7(b)}
\begin{figure}
	\centering
	\includegraphics[width=90mm,height=70mm]{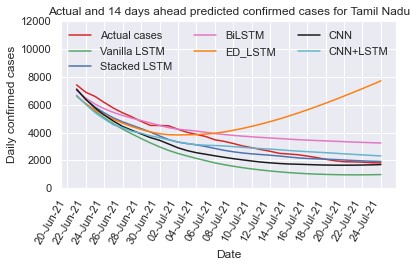}
	\vspace*{-9mm}
	\caption{Predicted and actual cases for Tamil Nadu for 14 days}
\end{figure}
\renewcommand{\thefigure}{7(c)}
\begin{figure}
	\centering
	\includegraphics[width=90mm,height=70mm]{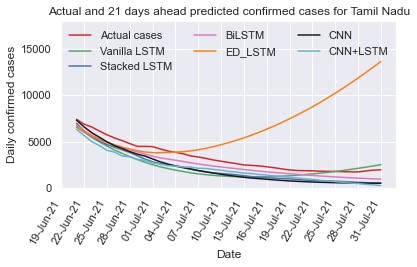}
	\vspace*{-9mm}
	\caption{Predicted and actual cases for Tamil Nadu for 21 days}
\end{figure}

\section{Conclusion}\label{con}
The COVID-19 outbreak is a potential threat due to its dynamical behaviour and more threatening in a country like India because it is very densely populated. The researchers are engaged in seeking new approaches to understand the COVID-19 dynamics that will overcome the limitation of existing epidemiological models. In this study, we designed the vanilla LSTM, stacked LSTM, ED-LSTM, Bi-LSTM, CNN, and hybrid CNN+LSTM model to capture the complex dynamical trends of COVID-19 spread and perform forecasting of the COVID-19 confirmed cases of 7, 14, 21 days for India and its four most affected states: Maharashtra, Kerala, Karnataka, and Tamil Nadu. The RMSE and MAPE errors on the testing data are computed to demonstrate the relative performance of the deep learning models.  The predicted COVID-19 confirmed cases of 7, 14, and 21 days for entire India and its states: Maharashtra, Kerala, Karnataka, and Tamil Nadu along with confidence intervals results shows that predicted daily confirmed cases by most of the models studied are very close to actual confirmed cases per day. The stacked LSTM and hybrid CNN+LSTM models perform better among the six models. These accurate predictions can help the governments to take decisions accordingly and create more infrastructures if required.
\section{Declaration of Competing Interest}\label{ci}
The authors declare that they have no known competing financial interests or personal relationships that could have appeared to influence the work reported in this paper.
\bibliographystyle{elsarticle-num}
\bibliography{mybib}
\end{document}